\documentclass[conference,a4paper]{IEEEtran}
\IEEEoverridecommandlockouts
\usepackage[noadjust]{cite} 
\usepackage[utf8]{inputenc}
\usepackage{hyperref}
\usepackage{graphicx}
\usepackage{booktabs}
\usepackage{amsmath}
\usepackage{amsthm}
\usepackage{xspace}
\usepackage{color}
\usepackage{xcolor,colortbl}
\usepackage{stfloats}
\usepackage{paralist}
\usepackage{subcaption}

\usepackage{hyperref}
\hypersetup{
colorlinks = true, 
linkcolor=black, 
citecolor=black, 
urlcolor=black 
}

\theoremstyle{definition}
\newtheorem{definition}{Definition}

\theoremstyle{definition}
\newtheorem{remark}{Remark}

\newcommand{\pp}{\ensuremath{\texttt{PP}}\xspace}
\newcommand{\ppW}{\ensuremath{\pp_{\weights}}\xspace}

\newcommand{\ppWPij}{\ensuremath{\pp_{\weights_i^j}}\xspace}

\newcommand{\timeout}{\ensuremath{\mathit{TO}}\xspace}
\newcommand{\dynObj}{\texttt{do}\xspace}
\newcommand{\dis}{\ensuremath{\mathit{dis}}\xspace}
\newcommand{\minDis}{\ensuremath{\mathit{minDis}}\xspace}
\newcommand{\comf}{\ensuremath{\mathit{comf}}\xspace}

\newcommand{\weights}{\ensuremath{\overline{w}}\xspace}
\newcommand{\weightsP}{\ensuremath{\overline{w}^\prime}\xspace}
\newcommand{\mo}{\ensuremath{\mathtt{MO}}\xspace}
\newcommand{\egopath}{\ensuremath{p_e}\xspace}
\newcommand{\egopathP}{\ensuremath{\egopath^\prime}\xspace}
\newcommand{\stp}{\ensuremath{\mathit{stp}}\xspace}

\newcommand{\cg}{\cellcolor{lightgray}}

\makeatletter
\newcommand{\linebreakand}{%
 \end{@IEEEauthorhalign}
 \hfill\mbox{}\par
 \mbox{}\hfill\begin{@IEEEauthorhalign}
}
\makeatother

\newcounter{researchquestionCount}
\newcommand{\researchquestion}[1]{\stepcounter{researchquestionCount}\vspace{5pt}\begin{compactitem}\item [\textbf{RQ\arabic{researchquestionCount}:}] \emph{#1}\end{compactitem}}

\begin{document}

\title{A Mutation-based Approach for Assessing Weight Coverage of a Path Planner\thanks{This work was supported, in part, by Science Foundation Ireland grant 13/RC/2094. T. Laurent is supported by an Irish Research Council grant (GOIPG/2017/1829). P. Arcaini and F. Ishikawa are supported by ERATO HASUO Metamathematics for Systems Design Project (No. JPMJER1603), JST. Funding Reference number: 10.13039/501100009024 ERATO.}}

\author{
\IEEEauthorblockN{Thomas Laurent}
\IEEEauthorblockA{\textit{Lero \& University College Dublin, Ireland} \\
\textit{National Institute of Informatics, Japan}\\
thomas.laurent@ucdconnect.ie}\\
\IEEEauthorblockN{Fuyuki Ishikawa}
\IEEEauthorblockA{\textit{National Institute of Informatics}\\
Tokyo, Japan\\
f-ishikawa@nii.ac.jp}
\and
\IEEEauthorblockN{Paolo Arcaini}
\IEEEauthorblockA{\textit{National Institute of Informatics}\\
Tokyo, Japan\\
arcaini@nii.ac.jp}\\
\IEEEauthorblockN{Anthony Ventresque}
\IEEEauthorblockA{\textit{Lero \& University College Dublin}\\ Dublin, Ireland\\
anthony.ventresque@ucd.ie}
}

\maketitle

\begin{abstract}
Autonomous cars are subjected to several different kind of inputs (other cars, road structure, etc.) and, therefore, testing the car under all possible conditions is impossible. To tackle this problem, scenario-based testing for automated driving defines categories of different scenarios that should be covered. Although this kind of coverage is a necessary condition, it still does not guarantee that \emph{any possible behaviour} of the autonomous car is tested. In this paper, we consider the path planner of an autonomous car that decides, at each timestep, the short-term path to follow in the next few seconds; such decision is done by using a weighted cost function that considers different aspects (safety, comfort, etc.). In order to assess whether all the possible decisions that can be taken by the path planner are covered by a given test suite $T$, we propose a mutation-based approach that mutates the weights of the cost function and then checks if at least one scenario of $T$ \emph{kills} the mutant. Preliminary experiments on a manually designed test suite show that some weights are easier to cover as they consider aspects that more likely occur in a scenario, and that more complicated scenarios (that generate more complex paths) are those that allow to cover more weights.
\end{abstract}


\section{Introduction}

Automated driving is a technology currently being intensely developed, that promises to impact our lives in many ways. Application for automated vehicles range from transport of goods (automated freight) to personal mobility, with offers such as Tesla's or Uber's. As promising as the technology is, great care must be taken in evaluating and validating such systems, to avoid tragic accidents happening~\cite{stewart_2018, levin_2018}.

Testing automated driving systems is critical for the satisfaction and safety of all stakeholders, but is also a very expensive operation. Therefore, it is essential to know when a system has been sufficiently tested. This is the question we focus on in this paper.

\begin{figure}[!tb]
\centering
\includegraphics[width=0.8\columnwidth, keepaspectratio]{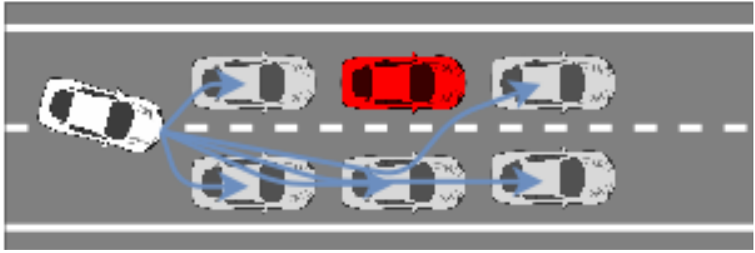}
\caption{Illustration of the path planner.}
\label{fig:pathPlanner}
\end{figure}

An autonomous driving system can be seen as a set of components that sense the environment of the vehicle, choose a path given an itinerary, and implement this path into concrete actions performed by actuators. In this paper, we consider a path planner component provided by our industry partner, which computes the best trajectory for the vehicle given a target destination. 
At every timestep, the path planner decides the {\it short-term path} that the car should follow in the next few seconds and the control commands that must be provided to implement it (such as acceleration and angle); in order to decide the next short-term path, an optimisation algorithm is employed. A set of short-term paths starting from the head of the car to a grid of points in front of the car are enumerated. This is shown in Fig.~\ref{fig:pathPlanner} where the white car is the ego car, the red car another, immobile, car. The translucent cars represent the possible future positions sampled by the path planner and the blue arrows the associated short-term paths.
Then, each short-term path is evaluated according to a cost function. The cost function considers different aspects, such as {\it safety}, {\it vehicle limitation}, {\it regulation compliance}, and {\it comfort}. Given the ranking of all the short-term paths, the one with the lowest cost is taken.

The path planner is tested using a simulator. The simulator takes as input a path planner and a scenario --a road configuration, a starting position, direction and speed for the automated car (or \emph{ego car}), and for other objects on the road-- and runs the path planner in this particular scenario, computing the path the ego car would take. Evaluating the output of the simulation, i.e., a path, is not done with a pass/fail oracle, as there is no comprehensive definition of what a valid path is. The presence or absence of a crash, for example, is not a good enough oracle: one can drive badly and not crash, or can take the best possible decision and still experience an unavoidable crash. Instead, we can define some metrics to capture some measures of interest (e.g., minimum distance of the ego car with all the other objects along the path); these metrics can be used to evaluate a path or compare paths computed by different path planners for a same scenario.

In testing, coverage criteria are used to evaluate the {\it quality} of a test suite $T$, i.e., if $T$ is sufficient for testing the System Under Test. For example, classical structural criteria check that the code has been covered sufficiently. In scenario-based testing~\cite{Menzel2018} for automated driving (as done with our simulator), main approaches aim at covering all different traffic situations (e.g., number and positions of the other cars), and manoeuvres done by the ego car~\cite{preprint_ITSC2019_Hauer}. To support this kind of testing, ontologies regarding driving behaviours, road topologies, environmental conditions, etc. have been devised (see the eight documents in~\cite{adsWaterlooWebsite}). Although such kind of coverage is also necessary for our path planner, it still does not guarantee coverage of all possible behaviours of the path planner.

Therefore, in this paper, we propose a definition of what it means to {\it sufficiently} test a path planner.
At a very high level, we want to check that all the {\it possible decisions} that can be taken by the path planner are observed in at least one test. This is difficult because, in general, we do not know which scenarios lead to some given decisions; this is why the coverage of scenario elements (as in~\cite{preprint_ITSC2019_Hauer}) may be not sufficient and so we need a different criterion. Moreover, we do not even have a proper characterisation of the different decisions, i.e., given two short-term paths computed by the path planner, we cannot say if they have been taken for the same reasons (i.e., for respecting the same aspects).

Although it is impossible to evaluate if all possible decisions are covered in a direct manner, we propose an indirect way of assessing this. The path planner can be seen as a weighted function of the different aspects listed before (i.e., safety, comfort, etc.). For each aspect, the path planner has one or more {\it weight}(s) set by the system designer. Such weights represent how important that aspect is in selecting a short-term path (we call this selection a {\it decision}). We claim that a minimal condition for testing the path planner is that each weight is shown to be ``relevant'' in at least one decision in one test. We say that a weight $w_i$ is {\it covered} by a scenario $s$ (i.e., that $w_i$ is ``relevant'' to at least a decision taken for $s$) if using a different weight value $w_i^\prime$ in the path planner, the computed path $p^\prime$ is {\it different} from the path $p$ computed with the original weight. Indeed, if changing the weight $w_i$ in all possible ways does not affect any decision, it means that the aspect considered by $w_i$ is irrelevant in that scenario.

Since trying all the possible alternative weights is infeasible (as their number is very large), we propose a mutation-based approach~\cite{papadakisMutationSurvey2019} that is able to estimate the {\it weight coverage} of a given test suite $T$. The approach consists in {\it mutating} a weight $w_i$ with a finite set of mutation operators; $w_i$ is considered covered by a test suite $T$ if, in at least one test in $T$, the path computed by the mutated path planner is {\it different} from the path computed by original path planner, according to a {\it mutation oracle}. We propose three mutation oracles that provide different guarantees in terms of coverage: the {\it path oracle} simply compares the two paths point-wise, the {\it safety oracle} compares the minimum distance of the ego car w.r.t. the other objects in the two paths, and the {\it comfort oracle} compares the smoothness of the paths. Note that some mutation oracles, as the path oracle, are more likely to say that two paths are different, while others, as the comfort oracle, are more demanding and require bigger differences in order to consider two paths different; in general, stronger mutation oracles provide stronger guarantees that two paths are significantly different~\cite{papadakisMutationSurvey2019}.

The rest of this paper is structured as follows: Section~\ref{sec:definitions} introduces some core definitions, and Section~\ref{sec:proposedApproach} presents our definition of weight coverage and a mutation-based approach to estimate it. Section~\ref{sec:experiments} presents some experiments we performed to evaluate our approach and discusses their results. Section~\ref{sec:discussion} further discusses some insights from the experiments, and Section~\ref{sec:threats} tackles some threats that could affect the validity of the proposed approach. Finally, Section~\ref{sec:related} reviews some related work, and Section~\ref{sec:conclusions} concludes the paper.

\section{Definitions}\label{sec:definitions}

In the following, we provide some definitions related to the path planner and its simulator.

\begin{definition}[Scenario]\label{def:scenario}
A \emph{scenario} $s$ describes the environment in which the ego car is operating. It is constituted of:
\begin{compactitem}
\item a map $M$ describing the road structure:
\item an initial position, speed, acceleration, and direction of the ego car;
\item a target destination of the ego car;
\item a set of static objects $\mathit{SO}$; each static object is characterised by its position in the map, and its size (length and width);
\item a set of dynamic objects $\mathit{DO}$; each dynamic object, in addition to position and size, is also characterised by its initial speed, acceleration, and direction;
\item a timeout \timeout; the scenario must be run until time \timeout.
\end{compactitem}
We will use the dot notation (e.g., $s.\timeout$) to access a particular field of a scenario.
\end{definition}

For the sake of conciseness, in the following, we consider static objects as dynamic objects having no velocity and no acceleration.

\begin{definition}[Path]\label{def:path}
A \emph{path} is a sequence of tuples $[(t_1,$ $l_1=(x_1, y_1),$ $d_1,$ $v_1,$ $a_1),$ $\ldots,$ $(t_n,$ $l_n=(x_n, y_n),$ $d_n,$ $v_n,$ $a_n)]$, where each tuple $i$ identifies a timestamp $t_i$, a location $l_i=(x_i, y_i)$ in the map, a direction $d_i$, a speed $v_i$, and an acceleration $a_i$. We use the dot notation to access tuple fields at a given time $t_i$ (e.g., $p.a_i$).
\end{definition}

Note that, for each dynamic object \dynObj of a scenario $s$, we can automatically compute its path $p$ up to the timeout $s.\timeout$\footnote{In the path planner simulator we are using, the behaviour of dynamic objects does not depend on the current situation, but only on the initial conditions specified in the scenario. For this reason, we can compute the path offline.}; we will write $\dynObj(s) = p$, where $p=[(t_1, l_1=(x_1, y_1), d_1, v_1, a_1),$ $\ldots,$ $(t_n, l_n=(x_n, y_n), d_n, v_n, a_n)]$ and $t_n = s.\timeout$.

\begin{definition}[Path Planner]\label{def:pathPlanner}
A \emph{path planner} \pp can be seen as a function that, given a scenario $s$, produces a path $p$ for the ego car up to simulation time $s.\timeout$, formally, $\pp(s) = p$. We name each pair of consecutive tuples $((t_i,$ $l_i=(x_i, y_i),$ $d_i,$ $v_i,$ $a_i),$ $(t_{i+1},$ $l_{i+1}=(x_{i+1}, y_{i+1}),$ $d_{i+1},$ $v_{i+1},$ $a_{i+1}))$ (with $i \in \{1,\ldots, n - 1\}$) as \emph{short-term path}: it corresponds to a \emph{decision} taken by the path planner.
\end{definition}

\subsection{Evaluation metrics (path quality metrics)}

Given a scenario $s$, we can define different metrics characterising the whole path computed by the path planner. In the following, let $\egopath = \pp(s)$ be the path computed by the path planner for the ego car, and $p_1 = \dynObj_{1}(s)$, \ldots, $p_m = \dynObj_{m}(s)$ be the paths of the dynamic objects $\dynObj_{1} \ldots \dynObj_{m}$.

\paragraph{Safety metric}
The first metric provides a {\it quantitative evaluation} of how safe the chosen path is. It is defined in terms of minimum distance between the ego car and any other object along the path as follows
\[\minDis(\egopath, \{p_1, \ldots, p_m\}) = \min_{\substack{i \in \{1 \ldots n\}\\j \in \{1 \ldots m\}}} \dis(\egopath.l_i, p_j.l_i)\]
where \dis is the Euclidean distance between two points.

\paragraph{Comfort metric} This metric assesses how comfortable the path has been for the driver. It is defined as maximum acceleration along the path:
\[\comf(\egopath) = \max_{i \in \{1 \ldots n\}} |\egopath.a_i|\]
Note that other comfort metrics could be defined in terms of, e.g., maximum torque or maximum lateral acceleration.

\section{Proposed approach}\label{sec:proposedApproach}

In this paper, we are interested in defining {\it sufficiency} criteria for path planner testing. A path planner can take different decisions on the base of the different environmental and driving conditions in which it is operating; we would like to check that all these possible decisions that can be taken by the path planner are observed in at least one test. However, we do not precisely know which scenarios cause a particular decision. Moreover, we cannot even characterise all the possible decisions taken by the path planner, i.e., given two decisions (two short-term paths computed by the path planner) we do not know if they have been selected for the same reason. However, we can exploit the architecture of the particular path planner under test in order to create a {\it proxy} for these decisions. In this section, we describe how we propose to do this using a mutation-based approach.

\subsection{Path planner under test}\label{sec:ppUnderTest}

The path planner provided by our industrial partner works as follows. At each timestep, it chooses which short-term path to follow in the next time period (see Def.~\ref{def:pathPlanner}). In order to do this, it enumerates a set of possible short-term paths, and scores them using a {\it weighted} cost function that considers different aspects:
\begin{itemize}
\item {\it Safety}: no collision with moving or static objects must happen and safety distances must be respected;
\item {\it Vehicle Limitation}: actions that cannot be achieved by the car must be avoided (e.g., no impossible steering can be required to follow a path);
\item {\it Compliance}: the car should respect road regulations as much as possible;
\item {\it Comfort}: the path should be as comfortable as possible for the passenger, avoiding too much forward and/or lateral acceleration.
\end{itemize}
In particular, the cost function uses these weights $W$ for the different aspects:
\begin{itemize}
\item
$w_1$: a factor that is multiplied with the maximum lateral acceleration along the short-term path;
\item
$w_2$: a constant that is added to the total cost if the maximum lateral acceleration is over a given threshold;
\item
$w_3$: a constant that is added to the total cost if the speed is over a given speed limit;
\item 
$w_4$: a constant that is added to total cost if the maximum acceleration along the short-term path is over a certain threshold;
\item 
$w_5$: a constant that is added to the total cost if the maximum deceleration along the short-term path is over a given threshold;
\item 
$w_6$: a constant that is added to the total cost if the curvature along the short-term path is over a given threshold.
\end{itemize}
As such, the different decisions that the system can take are tightly dictated by these weights. Note that weights $w_1$, $w_2$, $w_3$, $w_4$, and $w_5$ are related to the safety aspect; weights $w_1$, $w_2$, and $w_6$, are related to the comfort aspect; weight $w_4$ is related to the compliance aspect; finally, $w_2$, $w_4$, and $w_6$ are related to the vehicle limitation aspect. A weight can be associated with more than one aspect, e.g., $w_1$ is associated to the safety and comfort aspects, and $w_2$ to all the aspects.

Our industrial partner provided us with a version of the path planner that has been calibrated with a satisfactory set of weight values \weights for $W$. In the following, we identify with \ppW the path planner configured with weight values \weights.

\subsection{Weight coverage}\label{sec:weightCov}

Since the weights are strictly bound to the aspects that are considered in the decisions, we propose to map the coverage of the possible decisions with the coverage of the weights used to make the decisions.

Therefore, in this section, we propose a way to assess whether a weight is {\it involved} in a decision and, in section~\ref{sec:mutationBaseAppr}, a technique for measuring the sufficiency of a given test suite $T$ in testing the weights \weights.

\begin{definition}[Weight coverage criterion]\label{def:weightCov}
Given a path planner \ppW with weights $\weights = \{w_1, \ldots, w_k\}$, a test scenario $s$ \emph{covers} a weight $w_i \in \weights$ w.r.t. a metric $M$, if there exists a weight $w_i^\prime \neq w_i$ such that $M(\ppW(s)) \neq M(\pp_{\weights^{\prime}_i}(s))$ with $\weights^{\prime}_i = \{w_1,$ $\ldots,$ $w_i^\prime,$ $\ldots,$ $w_k\}$.
\end{definition}

Intuitively, a test scenario $s$ covers a weight $w_i$ if, with another value of the weight, the path planner behaves differently according to metric $M$. A good test suite $T$ should then cover all weights $w_i$.

Note that weight coverage has similarities with the MC/DC coverage criterion~\cite{mcdcChilenski1994} for Boolean expressions in which each clause must be shown to {\it determine} the value of the global predicate in a test: given an assignment of truth values, a clause $C$ determines the value of the global predicate $P$ if flipping the value of $C$ changes the value of $P$. In our case, for each weight $w_i$, we want to have a test in which the aspect considered by $w_i$ has some influence on the final decision taken by the path planner; we want to show that by modifying the weight in some way we can also modify the decision.

\begin{remark}
The path planner, in order to decide the next short-term path in a scenario $s$, assigns a numerical cost to a set of possible short-term paths $\stp_1, \ldots, \stp_n$, using a cost function that depends on the weights \weights; then, it selects the candidate with the lowest cost. Changing a weight $w_i$ in $w_i^\prime$ will change the cost of a given short-term path $\stp_j$ from $c_j$ to $c_j^\prime = c_j + \Delta_j$. If the weight considers an aspect that is relevant for the scenario $s$, $\Delta_j$ will be different for the different short-term paths and so their ranking could be modified (and so the final decision). Instead, if weight $w_i$ considers an aspect that is irrelevant for the scenario $s$, the costs of all the possible short-term paths will be modified by a same value $\Delta$; therefore, the rank of the possible short-term paths will not be affected and the same short-term path (i.e., the one selected with the original weights) will be selected as final decision.
\end{remark}

\subsection{Mutation-based approximation of weight coverage}\label{sec:mutationBaseAppr}
As we can not exhaustively evaluate the weight coverage of a test suite $T$ (the weights having continuous values), we propose a {\it mutation-based approach} to estimate whether or not $T$ covers the different weights.

In the following, we describe the mutation operators we use to generate mutants, some oracles that we use to assess whether a test kills a mutant, and finally how we use these for estimating weight coverage.

\subsubsection{Mutation operators}\label{sec:mutationOperators}

In this work, we are only concerned with the coverage of the test suite $T$ w.r.t. each individual weight $w_i$. Thus, we propose a simple mutation operator: each mutant $\weightsP_i$ differs only in the value $w_i^\prime$ of a weight $w_i$, which is multiplied by a constant $K$, i.e., $w_i^\prime = K \cdot w_i$. In order to explore different ranges for each weight, we use the following values of $K$: 0, 0.5, 0.9, 1.1, 1.5, 2, 10. This leads to seven versions of the operator that we refer to as $\mathit{MOs} = \{\mo_1, \ldots, \mo_7\}$. These factors were chosen to sample the space of possible weight values. In particular, 0 and 10 show extreme changes, 0 completely cancelling the effect of a weight. The other values of $K$ let us explore the effect of different scales of change to the weight values.

In the following, we identify with \ppWPij the path planner obtained from \ppW by mutating weight $w_i$ with mutation operator $\mo_j$.

\begin{remark}
Note that our mutation operators are not meant to be related to some fault-classes as in classical mutation analysis, i.e., they are not meant to replicate some possible faults. They are used to artificially perturbate the path planner, such that it possibly takes different decisions due to the mutated weight. As future work, we could design more targeted mutation operators, based on system and domain knowledge.
\end{remark}

\subsubsection{Mutation oracles}\label{sec:mutationOracles}
In order to assess whether a mutant has been killed, we need to compare the paths computed by the original path planner \ppW and the mutated one \ppWPij. In the following, given a scenario $s$, let $\egopath = \ppW(s)$ and $\egopathP = \ppWPij(s)$ be two paths computed by the two path planners, and $p_1 = \dynObj_{1}(s)$, \ldots, $p_m = \dynObj_{m}(s)$ be the paths of the dynamic objects $\dynObj_{1} \ldots \dynObj_{m}$. The mutated path planner is considered {\it killed} by $s$ if $\egopath$ and $\egopathP$ are {\it sufficiently different}.

In order to assess this, we can use different mutation oracles that differ in the characteristics of the paths they consider (e.g., safety or comfort). We devised the following oracles, defined as predicate $\mathit{killed}$ that tells whether a scenario $s$ kills the path planner \ppWPij obtained by mutating weight $w_i$ with mutation operator $\mo_j$.

\begin{itemize}
\item[\textbf{Path Oracle (PO)}] Given a threshold $\theta_P$, the mutated path planner is considered killed if there is a timestep in which the difference in the position of the ego car in the two paths is greater than $\theta_P$, i.e., 
\[\begin{array}{l}
\mathit{killed_P}(s, w_i, \mo_j) =\\
(\exists i \in \{1, \ldots, n\} \colon \dis(\egopath.l_i, \egopathP.l_i) > \theta_P)
\end{array}\]
where \dis is the Euclidean distance.
\item[\textbf{Safety Oracle (SO)}] Given a threshold $\theta_S$, the mutated path planner is killed if the difference of the minimum distances (with the dynamic objects) of the two paths is greater than $\theta_S$, i.e.,
\[\begin{array}{l}\mathit{killed_S}(s, w_i, \mo_j) =\\\left(\left|\begin{array}{l}\minDis(\egopath, \{p_1, \ldots, p_m\}) -\\ \minDis(\egopathP, \{p_1, \ldots, p_m\})\end{array}\right| > \theta_S\right)\end{array}\]
\item[\textbf{Comfort Oracle (CO)}] Given a threshold $\theta_C$, the mutated path planner is killed if the difference of the comfort measure in the two paths is greater than $\theta_C$, i.e.,
\[\mathit{killed_C}(s, w_i, \mo_j) = (|\comf(\egopath) - \comf(\egopathP)| > \theta_C)\]
\end{itemize}

Thresholds $\theta_P$, $\theta_S$, and $\theta_C$ must be selected by the domain expert who can tune how much difference must be observed in order to declare a mutant killed.

The different measures assess different aspects of the system but are also more or less ``strict''. For example, the Path Oracle is the ``easiest'' to kill, and should be subsumed by the other oracles (for equivalent thresholds). This idea can be somehow related to the idea of {\it weak} and {\it strong} mutation testing~\cite{papadakisMutationSurvey2019} in classic software testing, where weak mutation measures changes in the internal state of the program caused by mutants, while strong mutation considers only changes to output.

\subsubsection{Estimating weight coverage}
We can now describe a way to estimate weight coverage. From an original valuation \weights of weights $W$ for the path planner \pp, we create mutants \weightsP by applying the mutation operators described before, by changing the value of each weight $w_i$ in \weights in turn, thereby obtaining mutated versions of the path planner \ppWPij. We then run $T$ against each \ppWPij and determine which mutants are killed by $T$ according to our different oracles. Finally, following our initial definition of {\it weight coverage} (see Def.~\ref{def:weightCov}), we estimate that a weight $w_i$ is covered w.r.t. a metric $M$ (with $M \in \{P, S, C\}$) if one of its mutants is killed in a scenario, i.e.,
\begin{equation}\label{eq:covered}
\begin{array}{l}\mathit{covered}(w_i, T, M) =\\(\exists s \in T, \exists \mo_j \in \mathit{MOs} \colon \mathit{killed}_M(s, w_i, \mo_j))\end{array}
\end{equation}

By Def.~\ref{def:weightCov}, if $\mathit{covered}(w_i, T, M)$ holds, then $w_i$ is also covered for the weight coverage criterion. If $\mathit{covered}(w_i, T, M)$ does not hold, we can {\it estimate} that weight coverage does not hold as well, assuming that the mutation operators $\mathit{MOs}$ are a good proxy of all the possible weight changes (see Sect.~\ref{sec:threats} for a more detailed discussion on this point). In the remainder of the text, we use the phrase \textit{weight coverage} for both the coverage and its mutation-based approximation.

\section{Experiments}\label{sec:experiments}

In order to evaluate the approach, we designed a test suite $T$ composed of 10 scenarios, whose description is reported in Table~\ref{tab:scenariosDescription}.
\begin{table*}[!tb]
\centering
\caption{\textsc{Description of the test suite scenarios}}
\label{tab:scenariosDescription}
\begin{tabular}{c|p{0.94\textwidth}}
\toprule
ID & Description\\
\midrule
$s_1$ & The ego car is proceeding on a lane and two dynamic objects cross the street closely in front of it.\\
$s_2$ & The ego car is proceeding on a lane, following a slowing dynamic object and with a faster dynamic object coming from behind.\\
$s_3$ & The ego car is proceeding on a lane and a dynamic object is proceeding in the different direction on a different lane.\\
$s_4$ & The ego car is proceeding on a lane, encounters a parked car, and overtakes it.\\
$s_5$ & Similar to $s_4$, but there is another car coming from the opposite direction. The ego car has enough time to overtake the parked car before the other car arrives.\\
$s_6$ & Similar to $s_5$, but the ego car must let the other car pass before overtaking the parked car (as there is not enough time before).\\
$s_7$ & At a crossing, the ego car must turn right, while a dynamic object crosses the intersection from the opposite direction. The ego car must let the object pass before turning.\\
$s_8$ & At a crossing, the ego car must turn right, while a dynamic object is approaching the intersection from right. The ego car must slow down and let the dynamic object pass.\\
$s_9$ & At a crossing, the ego turns right, and, just after the turn, it encounters a dynamic object coming against the flow of traffic in its target lane.\\
$s_{10}$ & The ego car is approaching from behind a dynamic object that is slowing down.\\
\bottomrule
\end{tabular}
\vspace{-0.5cm}
\end{table*}
Note that the path planner is designed to work in countries as Ireland and Japan that adopt the left-hand traffic practice. While designing the test suite, we tried to cover different kinds of manoeuvres (e.g., going straight, overtaking a parked car, turning at a crossroad, giving precedence to another car, etc.). All the scenarios have been designed manually, except for scenario $s_1$ that has been found using a search algorithm with the aim of having a dangerous situation.

Then, we mutated the six weights $W = \{w_1, \ldots, w_6\}$ of the original path planner (see Sect.~\ref{sec:ppUnderTest}) using the seven mutation operators $\mathit{MOs} = \{\mo_1, \ldots, \mo_7\}$ described in Sect.~\ref{sec:mutationOperators}. Therefore, in total we have $6 \times 7 = 42$ mutated versions of the path planner; as before, we identify with \ppWPij the path planner obtained from \ppW by mutating weight $w_i$ with mutation operator $\mo_j$.

We then ran the designed test suite $T$ on the original path planner \ppW and the 42 mutated versions \ppWPij; we collected all the produced paths and computed the mutation oracles as specified in Sect.~\ref{sec:mutationOracles}. For the experiments, thresholds $\theta_P$, $\theta_S$, and $\theta_C$ of the mutation oracles have been set to 0: in this way, it is easier to compare the killing strength of each oracle.

We evaluated the approach using four research questions.

\researchquestion{What is the weight coverage of the designed test suite?}

We are interested in assessing how much the designed test suite covers the path planner weights. Table~\ref{tab:weightCoverage} reports, for each weight $w_i$, its coverage (either {\tt T}rue or {\tt F}alse) according to the three mutation oracles (see Eq.~\ref{eq:covered}); we highlight in grey the covered cases.
\begin{table}[!tb]
\centering
\caption{\textsc{Weight coverage (T: covered, F: not covered)}}
\vspace{-0.5cm}
\label{tab:weightCoverage}
\begin{tabular}{cccc}
\toprule
\textbf{Weight} & \multicolumn{3}{c}{\textbf{Mutation oracle}}\\
\cline{2-4}
& \textit{\textbf{PO}} & \textit{\textbf{SO}} & \textit{\textbf{CO}}\\
\midrule
$w_1$
& \cg T & \cg T & \cg T \\
\hline
$w_2$
& \cg T & \cg T & \cg T \\
\hline
$w_3$
& \cg T & F & F \\
\hline
$w_4$ 
& \cg T & \cg T & \cg T \\
\hline
$w_5$ 
& \cg T & \cg T & \cg T \\
\hline
$w_6$ 
& \cg T & F & F \\
\bottomrule
\end{tabular}
\vspace{-0.425cm}
\end{table}
Weights $w_1$, $w_2$, $w_4$, and $w_5$ are covered by the test suite with all the mutation oracles; these weights are all related to (lateral) acceleration/deceleration. The fact that they are all covered means that the test suite contains tests in which the acceleration has some effect on the decision taken by the path planner. Moreover, they are covered not only with the path oracle (that is a weak criterion for declaring a mutant killed), but also with the safety and comfort oracles, that are more demanding: this means that the killed mutants change both the minimum distance with the other dynamic objects (considered in the safety oracle) and the maximum acceleration (considered in the comfort oracle).

Weights $w_3$ and $w_6$ (related to the speed limit and sudden change of direction), instead, are only covered by the path oracle. This means that, although the mutants can slightly change the taken path, they do not affect the minimum distance with the other cars and the maximum speed.

\researchquestion{What is the weight coverage provided by each single scenario?}

We want to conduct a deeper analysis on the coverage provided by each single scenario. Table~\ref{tab:weightCoverageByScen} reports, for the three mutation oracles, whether a given scenario covers a given weight.
\begin{table*}[!tb]
\caption{\textsc{Weight coverage by scenario (T: covered, F: not covered)}}
\label{tab:weightCoverageByScen}
\begin{subtable}[t]{.33\textwidth}
\centering
\caption{Mutation oracle PO}
\label{tab:weightCoverageByScenP}
\setlength{\tabcolsep}{1.5pt}
\begin{tabular}{cccccccc}
\toprule
\textbf{s} & \multicolumn{6}{c}{\textbf{Weight}}& \textbf{Count}\\
\cline{2-7}
& $w_1$ & $w_2$ & $w_3$ & $w_4$ & $w_5$ & $w_6$\\
\midrule
$s_1$ & \cg T & \cg T & \cg T & \cg T & F & F & 4/6\\
\hline
$s_2$ & F & \cg T & F & \cg T & F & \cg T & 3/6\\
\hline
$s_3$ & F & F & F & F & F & F & 0/6\\
\hline
$s_4$ & \cg T & \cg T & F & F & F & F & 2/6\\
\hline
$s_5$ & \cg T & \cg T & F & \cg T & \cg T & F & 4/6\\
\hline
$s_6$ & F & \cg T & F & \cg T & \cg T & F & 3/6\\
\hline
$s_7$ & \cg T & \cg T & F & \cg T & \cg T & F & 4/6\\
\hline
$s_8$ & \cg T & \cg T & F & \cg T & \cg T & F & 4/6\\
\hline
$s_9$ & \cg T & \cg T & F & \cg T & \cg T & F & 4/6\\
\hline
$s_{10}$ & \cg T & F & F & \cg T & \cg T & F & 3/6\\
\midrule
Count & 7/10 & 8/10 & 1/10 & 8/10 & 6/10 & 1/10\\
\bottomrule
\end{tabular}
\end{subtable}
\begin{subtable}[t]{.33\textwidth}
\centering
\caption{Mutation oracle SO}
\label{tab:weightCoverageByScenS}
\setlength{\tabcolsep}{1.5pt}
\begin{tabular}{cccccccc}
\toprule
\textbf{s} & \multicolumn{6}{c}{\textbf{Weight}}& \textbf{Count}\\
\cline{2-7}
& $w_1$ & $w_2$ & $w_3$ & $w_4$ & $w_5$ & $w_6$\\
\midrule
$s_1$ & F & \cg T & F & F & F & F & 1/6\\
\hline
$s_2$ & F & \cg T & F & F & F & F & 1/6\\
\hline
$s_3$ & F & F & F & F & F & F & 0/6\\
\hline
$s_4$ & \cg T & F & F & F & F & F & 1/6\\
\hline
$s_5$ & \cg T & \cg T & F & \cg T & \cg T & F & 4/6\\
\hline
$s_6$ & F & \cg T & F & F & \cg T & F & 2/6\\
\hline
$s_7$ & \cg T & \cg T & F & \cg T & \cg T & F & 4/6\\
\hline
$s_8$ & F & F & F & F & \cg T & F & 1/6\\
\hline
$s_9$ & \cg T & \cg T & F & F & F & F & 2/6\\
\hline
$s_{10}$ & \cg T & F & F & \cg T & \cg T & F & 3/6\\
\midrule
Count & 5/10 & 6/10 & 0/10 & 3/10 & 5/10 & 0/10\\
\bottomrule
\end{tabular}
\end{subtable}
\begin{subtable}[t]{.33\textwidth}
\centering
\caption{Mutation oracle CO}
\label{tab:weightCoverageByScenC}
\setlength{\tabcolsep}{1.5pt}
\begin{tabular}{cccccccc}
\toprule
\textbf{s} & \multicolumn{6}{c}{\textbf{Weight}}& \textbf{Count}\\
\cline{2-7}
& $w_1$ & $w_2$ & $w_3$ & $w_4$ & $w_5$ & $w_6$\\
\midrule
$s_1$ & F & F & F & F & F & F & 0/6\\
\hline
$s_2$ & F & \cg T & F & F & F & F & 1/6\\
\hline
$s_3$ & F & F & F & F & F & F & 0/6\\
\hline
$s_4$ & F & F & F & F & F & F & 0/6\\
\hline
$s_5$ & F & F & F & F & F & F & 0/6\\
\hline
$s_6$ & F & F & F & F & F & F & 0/6\\
\hline
$s_7$ & \cg T & \cg T & F & \cg T & F & F & 3/6\\
\hline
$s_8$& F & F & F & F & F & F & 0/6\\
\hline
$s_9$& F & F & F & F & F & F & 0/6\\
\hline
$s_{10}$ & F & F & F & F & \cg T & F & 1/6\\
\midrule
Count & 1/10 & 2/10 & 0/10 & 1/10 & 1/10 & 0/10\\
\bottomrule
\end{tabular}
\end{subtable}
\vspace{-0.425cm}
\end{table*}
Considering mutation oracle PO, we observe that scenarios $s_1$, $s_5$, $s_7$, $s_8$, and $s_9$ cover more than half of the weights (4/6); these scenarios are among the most complicated ones (see the description in Table~\ref{tab:scenariosDescription}), in which different aspects must be taken into consideration; this also partially holds for scenarios $s_2$, $s_6$, and $s_{10}$ that cover half of the weights. Scenario $s_3$ does not cover any weight, as it simply describes a situation in which the ego car is going straight, and not too many factors influence the decision of the path planner in this case.

Regarding the mutation oracle SO, in general, scenarios kill fewer weights than what done with the mutation oracle PO: this is expected, as SO subsumes PO (i.e., if the minimum distance is different, the path must be different as well, but not the other way round). However, scenarios $s_5$, $s_7$, and $s_{10}$ kill the same weights with the two oracles; this means that, for these scenarios, the mutants always lead to a different path in which the minimum distance with the dynamic objects is affected (either smaller or larger). Indeed, using the original path planner, the ego car gets quite close to the dynamic objects, and so it is reasonable that any change in the path affects also the minimum distance.

Regarding the mutation oracle CO, only three scenarios kill some weight. Scenario $s_7$ achieves the highest coverage, killing half of the weights; this is due to the fact that the change of the weights leads to either a greater maximum acceleration to cross before the incoming car, or a lower maximum acceleration to let the other car pass, depending on the mutants (see scenario description in Table~\ref{tab:scenariosDescription}).

\researchquestion{How many scenarios cover each weight?}

We now want to assess how easy it is to cover a weight; we assume that the more scenarios cover a weight, the easier it is to cover it. The last rows of the tables in Table~\ref{tab:weightCoverageByScen} report the count of scenarios covering a given weight.

Using the mutation oracle PO, we observe that $w_1$, $w_2$, and $w_4$ are the weights that are easier to cover. Indeed, they are all related to lateral/normal acceleration and very likely a decision of the path planner depends on the acceleration (and so a perturbation of the weights changes the computed path).

Instead, weight $w_3$ (related to the violation of the speed limit) is only covered by scenario $s_1$ in which the ego car is close to collision with two other dynamic objects. We further observe that, for $w_3$, only mutant $\pp_{\weights_3^0}$, in which the constraint on the speed limit is completely removed, is killed: in this way, the path planner can compute an even safer (and so different) path that avoids the dynamic objects faster.

Also weight $w_6$ (related to sudden change of direction) is only covered by a single scenario, namely $s_2$. In $s_2$ the ego car is approaching a slowing car and is followed by a fast car that is approaching its back: by relaxing the constraint on the sudden change of lane, the path planner can compute a different and safer path.

Observations similar to those done for mutation oracle PO can also be done for mutation oracle SO. We only observe that $w_3$ is no more covered by $s_1$: this means that, although the mutated path planner can compute a different path, the minimum distance to the other dynamic objects remains the same (the mutated path planner can simply exit from the dangerous situation faster, as the constraint on the speed limit has been relaxed). In the same way, scenario $s_2$ no longer covers weight $w_6$: the mutated path planner can avoid the dangerous situation with a more sudden action (because the weight is relaxed) but it reaches the same minimum distance as the original path planner.

As we already observed in RQ2, the comfort oracle CO is highly demanding and it is difficult to kill mutants with this oracle (it is only possible by obtaining a path with a different maximum acceleration). As expected, only the weights related to acceleration (i.e., $w_1$, $w_2$, $w_4$, and $w_5$) can be covered by at least one scenario.

\researchquestion{What is the weight coverage provided by each mutation operator?}

We are interested in assessing which mutation operators produce mutants that are easier to kill. Table~\ref{tab:weightCoverageByMutOp} reports, for the three mutation oracles, whether a given mutation operator (we report the constant $K$ used in the operator) produces a mutated path planner that is covered (for at least one scenario of the test suite $T$).
\begin{table*}[!tb]
\caption{\textsc{Weight coverage by mutation operator (T: covered, F: not covered)}}
\label{tab:weightCoverageByMutOp}
\begin{subtable}[t]{.33\textwidth}
\centering
\caption{Mutation oracle PO}
\label{tab:weightCoverageByByMutOpP}
\setlength{\tabcolsep}{1.5pt}
\begin{tabular}{cccccccc}
\toprule
$\mathbf{K}$ & \multicolumn{6}{c}{\textbf{Weight}}& \textbf{Count}\\
\cline{2-7}
& $w_1$ & $w_2$ & $w_3$ & $w_4$ & $w_5$ & $w_6$\\
\midrule
0 & \cg T & \cg T & \cg T & \cg T & \cg T & F & 5/6\\
\hline
0.5 & \cg T & F & F & \cg T & \cg T & F & 3/6\\
\hline
0.9 & F & F & F & \cg T & F & F & 1/6\\
\hline
1.1 & F & F & F & F & \cg T & F & 1/6\\
\hline
1.5 & F & \cg T & F & \cg T & \cg T & F & 3/6\\
\hline
2 & F & \cg T & F & \cg T & \cg T & F & 3/6\\
\hline
10 & \cg T & \cg T & F & \cg T & \cg T & \cg T & 5/6\\
\midrule
Count & 3/7 & 4/7 & 1/7 & 6/7 & 6/7 & 1/7\\
\bottomrule
\end{tabular}
\end{subtable}
\begin{subtable}[t]{.33\textwidth}
\centering
\caption{Mutation oracle SO}
\label{tab:weightCoverageByMutOpS}
\setlength{\tabcolsep}{1.5pt}
\begin{tabular}{cccccccc}
\toprule
$\mathbf{K}$ & \multicolumn{6}{c}{\textbf{Weight}}& \textbf{Count}\\
\cline{2-7}
& $w_1$ & $w_2$ & $w_3$ & $w_4$ & $w_5$ & $w_6$\\
\midrule
0 & \cg T & \cg T & F & \cg T & \cg T & F & 4/6\\
\hline
0.5 & F & F & F & \cg T & \cg T & F & 3/6\\
\hline
0.9 & F & F & F & \cg T & F & F & 1/6\\
\hline
1.1 & F & F & F & F & \cg T & F & 1/6\\
\hline
1.5 & F & \cg T & F & F & \cg T & F & 2/6\\
\hline
2 & F & \cg T & F & F & \cg T & F & 2/6\\
\hline
10 & \cg T & \cg T & F & \cg T & \cg T & F & 4/6\\
\midrule
Count & 2/7 & 4/7 & 0/7 & 4/7 & 6/7 & 0/7\\
\bottomrule
\end{tabular}
\end{subtable}
\begin{subtable}[t]{.33\textwidth}
\centering
\caption{Mutation oracle CO}
\label{tab:weightCoverageByMutOpC}
\setlength{\tabcolsep}{1.5pt}
\begin{tabular}{cccccccc}
\toprule
$\mathbf{K}$ & \multicolumn{6}{c}{\textbf{Weight}}& \textbf{Count}\\
\cline{2-7}
& $w_1$ & $w_2$ & $w_3$ & $w_4$ & $w_5$ & $w_6$\\
\midrule
0 & F & \cg T & F & \cg T & \cg T & F & 3/6\\
\hline
0.5 & F & F & F & F & F & F & 0/6\\
\hline
0.9 & F & F & F & F & F & F & 0/6\\
\hline
1.1 & F & F & F & F & F & F & 0/6\\
\hline
1.5 & F & F & F & F & F & F & 0/6\\
\hline
2 & F & \cg T & F & F & F & F & 1/6\\
\hline
10 & \cg T & \cg T & F & F & \cg T & F & 3/6\\
\midrule
Count & 1/7 & 3/7 & 0/7 & 1/7 & 2/7 & 0/7\\
\bottomrule
\end{tabular}
\end{subtable}
\vspace{-0.45cm}
\end{table*}
For all the mutation oracles, coverage is correlated with the degree of change of the weight: mutation operators that change the weight significantly (i.e., 0 and 10) are those that cover the most (5 out of 6 weights), while weaker mutation operators (i.e., 0.9 and 1.1) cover less. For mutation oracle CO, only mutants with $K$ equal to 0, 2, or 10, can lead to the coverage of at least one weight.

The results in Table~\ref{tab:weightCoverageByMutOp} also provide some insights on the weights themselves. Let's consider the results of the mutation oracle PO in Table~\ref{tab:weightCoverageByByMutOpP}. We observe that some weights such as $w_4$ and $w_5$ are covered with almost any mutation operator: this means that the weight is important in the decision making of the path planner and thus it is sensitive to small changes. On the other hand, if a system designer knew their test suite is strong, but a weight is not covered, this could show that the weight has no influence on the decision making, and could highlight a fault in the system or its design.

\section{Discussion}\label{sec:discussion}

We now provide more general observations about the proposed approach.

The first observation is related to the coverage of a weight. If a weight $w_i$ is never covered in a test suite, it could mean that either the test suite is not complete enough to cover $w_i$, or $w_i$ is never relevant in the decisions taken by the path planner. In the former case, we would just need to add some scenario trying to cover $w_i$; in the latter case, we would need to mark $w_i$ as an {\it infeasible} test requirement and we could report a problem in the path planner. However, detecting infeasible test requirements is in general undecidable.

Another observation is related to the {\it completeness} of the mutation-based approach. In order to approximate weight coverage of a weight $w_i$, we propose to use a set of seven mutants where $w_i$ is modified using seven constants of different scales. It could be that $w_i$ is covered by a test suite $T$ according to weight coverage (see Def.~\ref{def:weightCov}), but not using the mutation-based approximation. However, we believe that this does not affect the general conclusions of our experiments regarding the relations between the scenarios and the weights: it is unlikely that, given two scenarios that do not cover any weight with the mutation-based approach, there are some other changes of the weights (not considered by the mutants) that cover one scenario and not the other. Indeed, the path planner considered in this work uses a linear cost function. For a new value to change the result of a test when no mutant does, it would then need to induce a greater change than the mutants, which have already been designed to induce significant changes to the weights. Still, if such a case occurred, one could ponder the significance of such a coverage: does a scenario meaningfully cover a weight, if for the decision of the path planner in this scenario to change one must introduce massive change to the weight?

A final observation is related to the mutation oracles. We can note that, although the path oracle is a very weak criterion, it is still useful to decide whether a scenario should be kept in the test suite: if a scenario cannot kill any weight even with regards to the path oracle, it means that it is not challenging the path planner at all and should be removed from the test suite (as scenario $s_3$. See Table~\ref{tab:weightCoverageByScenP}).

\section{Threats to validity}\label{sec:threats}

We identify these threats to the validity of the approach.

A threat to external validity~\cite{Wohlin2012} is that the approach may not be generalizable to other systems. As this is a project driven by a collaboration with an industrial partner, the solution has the risk to be too domain specific. First of all, we want to point out that, in some cases, a solution to a given problem is necessarily domain-specific and trying to achieve generability could also be counterproductive~\cite{BriandBNPS17}. Moreover, we still believe that the approach could be applied to other systems similar to the path planner, i.e., systems solving some optimization problems using some weights to consider different aspects. As future work, we plan to evaluate whether the approach is generalizable to a broader class of systems. 

A threat to internal validity~\cite{Wohlin2012} could be that our mutation-based approach could be faulty and so the obtained results would be not meaningful. In order to mitigate this threat, we checked that the mutated scenarios are syntactically correct and that they are parsed correctly by the path planner simulator; moreover, we assessed that the mutation oracles are implemented correctly by verifying that some known relations between them hold: for example, if the path oracle is 0, the other two oracles must be 0 as well.

\section{Related work}\label{sec:related}

In this section we review some related work concerning testing of, and testing criteria for automated driving systems, as well as non-conventional applications of mutation analysis.

Testing of autonomous driving systems is a complex issue that includes many challenges, as highlighted by Koopman and Wagner in~\cite{2016-01-0128}. In~\cite{Wachenfeld2016}, Wachenfeld and Winner show that it is infeasible to test autonomous driving systems only using real life test drives. Indeed, they show that, according to German highway driving data, one would have to drive 6.61 billion kilometers in order to encounter some fatal scenarios, i.e., the scenarios that should be most critically tested. Zhao and Peng~\cite{DBLP:journals/corr/ZhaoP17aa} and Helmer et al.~\cite{Helmer2015} arrive at similar conclusions, stating that billions of kilometers should be driven to achieve sufficient testing guarantees.

As such, many works~\cite{Menzel2018,Roesener2016,Gelder2017,Helmer2015} focus on using simulation and particular scenarios to test autonomous driving systems, this is the situation we are in in the context of this work. In this context, the question of test sufficiency, or of a test stopping criterion becomes essential. Indeed, as Hauer at al. remark in~\cite{preprint_ITSC2019_Hauer}, ``One can always come up with another scenario type as well as with instances of those types that are different from the types and instances used before'', which means that we need a criterion to know when our test data has covered all plausible situations. Our work not only focuses on a test ending criterion but on a more general testing criterion that lets us evaluate how much of the system's decision space a test suite covers, rather than how much of the possible scenarios have been covered, as different scenarios could lead to the same decisions.

Mutation analysis has been applied to diverse domains~\cite{papadakisMutationSurvey2019}, and recently to deep neural networks (DNNs). DNNs have the same characteristic as the path planner, in that their behavior is governed by computed numerical values, rather than logical branches, and that their correctness is evaluated by some metrics (e.g., accuracy) rather than with pass/fail tests. A mutation analysis method for DNNs has been proposed that considers mutations on training data, training code, and trained models~\cite{ISSRE18-DeepMutation}. The mutation score evaluates whether each mutation changes correct classification into misclassification in some test data. Our proposal works with the more complex situation of path planner. Although the mutation targets (weights) are also continuous values, we deal with complex oracles and multiple evaluation criteria, instead of the binary problem of misclassification.

\section{Conclusions}\label{sec:conclusions}

In this paper, we proposed a mutation-based approach for assessing whether all the possible {\it decisions} that can be taken by the path planner of an autonomous car are covered in a test suite (each test is a scenario). The path planner we consider makes decisions by using a weighted function of different aspects (safety, comfort, etc.). The approach consists in mutating the weights and checking whether the test suite is able to kill the mutant. The approach has been experimented on a manually designed test suite; we observed that some weights are easier to cover as they consider aspects that occur more often in a scenario. Moreover, more complicated scenarios that generate more complex paths are those that allow coverage of more weights. We believe that these preliminary results confirm our intuition that the proposed coverage criterion is reasonable. However, more rigorous and systematic evaluation is needed: as future work, we plan to perform a wider set of experiments using different test suites, automatically generated and manually designed. Moreover, we plan to assess whether weight coverage correlates with good fault detection.

Finally, we believe that the proposed approach is not only applicable to path planners, but to any optimisation program that relies on a weighted function; as future work, we plan to give a more general definition of the weight coverage criterion, and experiment it on a wider class of systems.

\bibliographystyle{IEEEtran}
\bibliography{apsec2019}

\end{document}